# *Dispersion Synthesis with Multi-Ordered Metatronic Filters*

Yue Li[1,2] Iñigo Liberal[2] and Nader Engheta[2,*]

[1]Department of Electrical Engineering, Tsinghua University, Beijing 100084, China
[2] Department of Electrical and Systems Engineering, University of Pennsylvania, Philadelphia, Pennsylvania 19104, USA

**Abstract**

We propose the synthesis of frequency dispersion of layered structures based on the design of multi-ordered optical filters using nanocircuit concepts. Following the well known insertion loss method commonly employed in the design of electronic and microwave filters, here we theoretically show how we can tailor optical dispersion as we carry out the design of several low-pass, high-pass, band-pass and band-stop filters of different order with a (maximally flat) Butterworth response. We numerically demonstrate that these filters can be designed by combining metasurfaces made of one or two materials acting as optical lumped elements, and, hence, leading to simple, easy to apply, design rules. The theoretical results based on this circuital approach are validated with full-wave numerical simulations. The results presented here can be extended to virtually any frequency dispersion synthesis, filter design procedure and/or functionality, thus opening up exciting possibilities in the design of composite materials with on-demand dispersion and high-performance and compact optical filters using one or two materials.

[*] To whom correspondence should be addressed: Email: engheta@ee.upenn.edu

Filters are indispensable components in both microwave and optical systems generally aimed to attenuate signals located in unwanted frequency ranges. However, although filters are ubiquitous across the electromagnetic spectrum, different design strategies are utilized as function of the wavelength of operation. For instance, at the infrared and optical regimes, filters are constructed by using resonators such as optical gratings [1,2], microrings [3–6], defects in photonic crystals [7–9], and/or multi-coupled waveguides [10,11]. Since these devices are intrinsically based on interference phenomena, the dimensions of the filters are relatively large as compared to the wavelength of operation [1–11]. Contrarily, at the radio frequency (RF) and microwave regimes, filters are usually built by using lumped circuit elements [12,13], e.g. inductors, capacitors, and resistors. In this manner, the dimensions of the filters are in a deeply subwavelength scale [12,13]. Therefore, a question may naturally arise: is it possible to design optical filters based on the same concepts of lumped subwavelength components as it is customary for electronic and microwave filters? If so, this methodology may provide a useful approach for design of materials with desired frequency dispersion, since the constituent building blocks would be much smaller than operating wavelengths of interest.

In this regard, optical metatronics [14,15], i.e., metamaterial-inspired optical nanocircuitry, appears as an excellent candidate to transfer into the optical domain the tools and techniques aimed to the design of electronic and microwave higher-order filters. In essence, metatronic circuits control the flow of displacement current and, to this end, nanostructures play the role of lumped circuit elements [14]. As in electronics, this modularization approach enables to construct complex circuits by assembling individual and lumped circuit elements [16–18]. This strategy has already been exploited in the design of nanoantenna systems [19–23], metasurfaces [24] and computational devices [25]. Optical metatronic circuits have also been successfully employed in the design [26,27] and experimental demonstration [16–18] of optical nanofilters. These preliminary studies serve to explain the underlying physics of metatronic filters, as well as to illustrate how the design techniques for electronic circuits can be applied to optical components. However, practical applications require higher-order filters with flatter passbands and more abrupt transitions to the stopbands, which would also enable synthesis of composite structures with on-demand desired frequency dispersion. Here, we present the design and full-wave analysis of higher-order metatronic filters formed by collections of properly designed metasurfaces made of one or two materials. Specifically, we make use of the insertion loss method [13] in order to design higher-order optical filters with a (maximally flat) Butterworth response. We present designs of low-pass, high-pass, band-pass and band-stop filters using stacked metasurfaces, demonstrating that such components can be readily designed as the combination of simple lumped components consisting of electrically thin slabs of one or two materials.

It is worth remarking that previous works have already addressed the design of spatial and frequency filters based on stacked planar layers at microwave [28,29], infrared [30–34] and visible [34–36] frequencies, including the design of higher-order filters [37,38]. These works are based on early efforts in frequency selective surfaces [28] (FSS), where the devices are often designed on the basis of equivalent circuit models of FSS structures [39]. Typically, analytical or numerical methods are employed to solve the many-body scattering problem of the elements composing a FSS [29], and, then, once the solution is known, the collective response is encapsulated into an equivalent or averaged grid impedance [39].

By contrast, metatronic circuits do not represent equivalent circuits. On the contrary, they locally operate as *actual* circuits controlling the flow of the displacement current. The advantages of this approach are twofold: First, metatronic circuits are inherently simple, i.e., instead of solving many-body scattering problems in order to define an equivalent impedance, metatronic circuit elements are characterized by simple geometries with known impedance expressions (albeit occasionally approximate), physically connected to the flow of displacement current. Second, since elements of metatronic circuits actually and locally function as lumped circuit elements, they can be combined via series and parallel connections to construct more complex circuits, as if they were in an electronic circuit board. This fact facilitates the transfer to the optical domain of the tools aimed to the design of electronic circuits. For example, in the present work we will show that, once a metatronic optical filter has been designed, it can be easily converted into other classes of filters by using a known set of impedance transformations.

*Metatronic circuit elements* **-** Before delving into the details of the dispersion synthesis and design process we start by introducing the individual circuit elements, e.g., capacitors and inductors, composing first-order optical filters. To this end, we consider a material slab (of infinite size along the x- and y- axes) with thickness $a$ much smaller that the wavelength of operation $a \ll \lambda$. We assume that the slab is illuminated under normal incidence by an incident plane-wave with the electric field polarized along $\hat{y}$, i.e., the electric field is parallel to the surface of the slab. The free-space propagation[*] of this wave can be modeled as a transmission line with intrinsic impedance equal to the free-space medium intrinsic impedance $Z_0 = \sqrt{\mu_0/\epsilon_0}$,

---

[*] Although here for the sake of simplicity we assume the space between the slabs is free space, one can easily generalize this method to have dielectric spacers.

whereas, for this polarization, the thin slabs behave as a shunt circuit elements, thus operating as first-order optical nanofilters[24,25, 26,27] (see Fig. 1). For instance, as schematically depicted in Fig. 1(a) (green color), when the slab is composed of a dielectric (nonmagnetic) material of relative positive permittivity $\varepsilon_d$, it can be modeled as a shunt capacitor (i.e., a first-order low-pass filter) whose capacitance can be simply written as follows

$$C_{\text{slab}} = a\, \varepsilon_d \varepsilon_0 \qquad (1)$$

On the other hand, as shown in Fig. 1(b) (yellow color), a plasmonic slab with negative and dispersive relative permittivity $\varepsilon_m(\omega)$ behaves as a shunt inductor (i.e., a first-order high-pass filter), whose inductance is given by [24,25]

$$L_{\text{slab}} = -\frac{1}{a\omega^2\, \varepsilon_m(\omega)\varepsilon_0} \qquad (2)$$

For the sake of simplicity, we will consider plasmonic slabs characterized by an ideally lossless Drude model $\varepsilon_m(\omega) = 1 - \omega_p^2/\omega^2$, where $\omega_p$ stands for the plasma frequency. In general, this material model results in an intrinsically dispersive effective inductance, although this inductance value is approximately nondispersive at frequencies where $\varepsilon_m(\omega) \sim -\omega_p^2/\omega^2$, so that $L_{\text{slab}} \sim \omega_p^2/(a\varepsilon_0)$, a constant value. We remark that these very simple impedance expressions are only adopted and utilized for the design procedure at the circuit level. In fact, all results and designs will be subsequently validated by means of full-wave numerical simulations in which the size and dispersion characteristics of the slabs are fully taken into account. For the sake of clarity and simplicity in describing the main concept, in the present work we assume the materials are lossless, although one can easily add loss to the dispersion characteristics of each constituent materials.

One of the main advantages of the metatronic modularization of the material slabs as circuit elements is that it enables us to readily combine them in series and parallel connections [26,27]. As a byproduct, we can straightforwardly develop first-order band-pass and band-stop filters. On the one hand, if the plasmonic and non-plasmonic slabs are arranged alternatively along the z-axis, the interface between the two slabs is parallel to the electric field (see Fig. 1(c)). Consequently, the pair of slabs operates as the parallel connection of capacitor and inductor elements (i.e., a first-order band-pass filter). On the other hand, if the slabs are arranged alternatively along the y-axis, the narrow interface between the two strips is perpendicular to the electric field (see Fig. 1(d)). In this manner, the slabs are operating as series capacitor and inductor (i.e., a first-order band-stop filter).

Subsequently, these first-order circuit elements can be used as the building blocks of more complex higher-order optical filters. In principle, any electronic filter design procedure could be implemented by using these metatronic circuit elements. Here, we focus in the design of optical filters with a (maximally flat) Butterworth response by using the insertion loss method. However, this must be considered only as a particular example of the many design procedures that could be inherited from electronic circuit design in order to synthesize layered structures with given desired frequency dispersion. For instance, very similar techniques could be employed in the design of filters with the sharpest possible transition from the pass- to the stop- bands (Chebyshev response) or a better phase response (linear phase filter design).

In short, the insertion loss method describes the insertion loss of a filter as a combination of polynomials in $\omega^2$ that give rise to the optimal response according to some specific design criteria for a given filter order $N$ (e.g., Butterworth response for the flattest possible passband). In general, the order $N$ of the filter corresponds to the number of circuit units composing it. Consequently, the impedance values of the circuit configurations that adjust to this optimal polynomial frequency response can be derived by using analytical or numerical methods.

*Higher-order low-pass filter design* **-** To begin with, let us consider the design of a low-pass optical filter whose 3-dB cutoff angular frequency equals $\omega_{3dB}$. In this case, the optimal response is obtained by a succession of shunt capacitors and series inductors (c.f., Figs. 2(a) and 2(b) for the circuit diagrams of 2nd- and 3rd-orders low-pass filters, respectively). The associated optimal impedance values can be derived by fitting the prescribed frequency response, and their tabulated values are available in the literature in filter design [12]. In particular, the values of the n-th (capacitor or inductor) element of the aforementioned low-pass filters can be written as follows [12]:

$$C_n^{lp} = g_n/(Z_0 \omega_{3dB}) \text{ for } n = 1, 3, \ldots \qquad (3)$$

$$L_n^{lp} = g_n Z_0/\omega_{3dB} \quad \text{ for } n = 2, 4, \ldots \qquad (4)$$

where $g_n$ are tabulated coefficients that have been determined numerically and are available in the literature. For the sake of completeness and easy reference here, the coefficients for filters up to the third order are included in the first column of Table I. The other higher order values can be found in the literature [12].

Once the required impedance values are known, it is in principle possible to find a physical implementation for the individual circuit elements composing the filter. However, due to the polarization of the electric field of the incident wave, only shunt components can be implemented by stacking the material slabs as shown in Fig. 1. In Ref [25], a method to overcome this challenge using the epsilon-near-zero (ENZ) and mu-near-zero (MNZ) structures was studied. However, this difficulty can also be circumvented by using impedance transformations from transmission line theory [13,37,38]. Specifically, a shunt capacitor $C$ connected to λ/4 a transmission line provides the same input impedance as a series inductor $L = Z_0^2 C$. In this manner, the $L_n^{lp}$ series inductors can be replaced by a λ/4 transmission line section terminated in a capacitor of value $C_n^{lp} = \frac{L_n}{Z_0^2} = \frac{g_n}{Z_0 \omega_{3dB}}$. Note that for the 3$^{rd}$ and higher-order filters an additional λ/4 a transmission line section must be included right after this element in order to ensure that the next element behaves as a shunt capacitor. Naturally, this transformation is only exact at the frequency at which the electrical length of the transmission line section equals λ/4. In our case, we set the length of the section as $\lambda_{3dB}/4$, (a quarter wavelength at the cut-off frequency $\omega_{3dB}$), and it will be shown via numerical simulations that this choice results in a satisfactory performance over the entire frequency band. In summary, a low-pass filter of order $N$ can be constructed as a series of $N$ shunt capacitors with capacitance $C_n = g_n/(Z_0 \omega_{3dB})$ separated by $\lambda_{3dB}/4$ transmission line sections.

*Filter transformations* - Interestingly, applying frequency transformations to the insertion loss method enable one to design other types of filters, such as high-pass, band-pass, and band-stop filters, as described in Ref. [12]. We demonstrate further that the modularization tools provided by optical metatronics empower us to apply the same transformations and design procedures. To begin with, the substitution $\omega \rightarrow -\omega_{3dB}/\omega$ in the design process transforms a low-pass into a *high-pass* filter with the same 3dB cutoff frequency. Moreover, by applying this frequency transformation it can be found that the impedance transformation required to obtain a high-pass filter consists of replacing each $C_n^{lp}$ capacitor by an inductor $L_n^{hp}$ whose inductance is given by [12]:

$$L_n^{\text{hp}} = \frac{Z_0}{g_n \omega_{3dB}} \quad (5)$$

Similarly, the frequency substitution $\omega \rightarrow \Delta^{-1}(\frac{\omega}{\omega_0} - \frac{\omega_0}{\omega})$ transforms a low-pass response into a *band-pass* response with fractional bandwidth $\Delta = (\omega_2 - \omega_1)/\omega_0$ centered at the frequency $\omega_0 = \sqrt{\omega_1 \omega_2}$, where $\omega_1$ and $\omega_2$ are the prescribed frequency edges of the pass-band filter. Additionally, it can be proven that the required impedance transformation consists of replacing each capacitor $C_n^{lp}$ by a parallel LC circuit unit (see Fig. 3(c)) with values [12]

$$C_n^{bp} = \frac{g_n}{Z_0 \omega_0 \Delta} \quad (6)$$

$$L_n^{bp} = \frac{Z_0 \Delta}{g_n \omega_0} \quad (7)$$

Inversely, the frequency substitution $\omega \rightarrow -\Delta(\frac{\omega}{\omega_0} - \frac{\omega_0}{\omega})^{-1}$ transforms a low-pass frequency response into a *band-stop* response with fractional bandwidth $\Delta$ centered at $\omega_0$. In this case, the impedance transformation required to obtain a band-stop filter consists of substituting the capacitors $C_n^{lp}$ by parallel LC circuit unis (see Fig. 3(d)) with values [12]:

$$C_n^{bs} = \frac{g_n \Delta}{Z_0 \omega_0} \quad (8)$$

$$L_n^{bs} = \frac{Z_0}{g_n \omega_0 \Delta} \quad (9)$$

*Metatronic implementation* – The metatronic implementation of the aforementioned circuit configurations is catalyzed by the simplicity of the impedance expressions, equations (1) and (2), of the slabs acting as lumped elements. In fact, simple design rules of the thicknesses and materials required for the slabs to compose the filters can be readily derived on the basis of those expressions. To begin with, it is clear by comparing Figs. 3(a) and 1(a) that a low-pass filter of order $N$ can be implemented by stacking $N$ dielectric slabs separated by $\lambda_{3dB}/4$. Specifically, the thickness of the n-th slab can be determined by introducing equation (1) into (3), leading to

$$a_n^{lp} = \frac{g_n}{Z_0 \omega_{3dB} \varepsilon_d \varepsilon_0} \tag{10}$$

Similarly, a high-pass filter can be implemented by stacking plasmonic slabs (see Figs. 3(b) and 1(b)). In this case, the thickness of each slab is found by introducing equation (2) into (5), leading to

$$a_n^{hp} = -\frac{g_n}{Z_0 \omega_{3dB} \varepsilon_m(\omega_{3dB}) \varepsilon_0} \tag{11}$$

Next, a band-pass filter of order $N$ requires parallel LC circuit units (Fig. 3(c)) that can be implemented as pairs of adjacent dielectric and plasmonic slabs (Fig. 1(c)). Both the total thickness $a_n$ and thickness ratio $r_n$ of the slab (as described in Fig. 3(g)) are determined by the central frequency $\omega_0$ and fractional bandwidth $\Delta$, as well as the available materials, i.e., the permittivities of the dielectric $\epsilon_d$ and plasmonic $\epsilon_m$ slabs. By introducing the parallel connection of equations (1) and (2) into equations (6) and (7), we have:

$$r_n^{bp} = \frac{\varepsilon_d}{\varepsilon_d - \varepsilon_m(\omega_0)} \tag{12}$$

$$a_n^{bp} = g_n \frac{\varepsilon_m(\omega_0) - \varepsilon_d}{\Delta Z_0 \omega_0 \varepsilon_0 \varepsilon_d \varepsilon_m(\omega_0)} \tag{13}$$

To finalize, a band-stop filter of order $N$ requires series LC circuit units (see Fig. 3(d)) that can be implemented by using inhomogeneous slabs constructed by strips of dielectric and plasmonic materials (see Fig. 1(d)). The thickness of the slab $a_n$ and the height ratio $r_n$ (as described in Fig. 3(h)) can be determined by inserting the series connection of equations (1) and (2) into equations (8) and (9). This exercise leads to the following expressions

$$r_n^{bs} = \frac{\varepsilon_m(\omega_0)}{\varepsilon_m(\omega_0) - \varepsilon_d} \tag{14}$$

$$a_n^{\text{bs}} = \frac{g_n \Delta}{Z_0 \omega_0 \varepsilon_0 (\varepsilon_d - \varepsilon_m(\omega_0))} \tag{15}$$

TABLE I. Tabulated coefficients and dimensions for the physical implementation of different higher-order filters. For values of $\varepsilon_d$ and $\varepsilon_m$ for each case, see the text.

| Filter order | $g_{n\ (\text{form [12]})}$ | Low-pass: $f_{3dB}$=400 THz | High-pass: $f_{3dB}$=400 THz | Band-pass: $f_1$=410 THz, $f_2$=610 THz | Band-stop: $f_1$=190 THz, $f_2$=210 THz |
|---|---|---|---|---|---|
| 1st | $g_1$=2.000 | $a_1$=24.0 nm | $a_1$=9.92 nm | $a_1$=14.2 nm, $r_1$=0.34 | $a_1$=3.18 nm, $r_1$=0.34 |
| 2nd | $g_1$=1.414 | $a_1$=16.8 nm | $a_1$=7.02 nm | $a_1$=10.0 nm, $r_1$=0.34 | $a_1$=2.26 nm, $r_1$=0.34 |
|  | $g_2$=1.414 | $a_2$=16.8 nm | $a_2$=7.02 nm | $a_2$=10.0 nm, $r_2$=0.34 | $a_2$=2.26 nm, $r_2$=0.34 |
| 3rd | $g_1$=1.000 | $a_1$=12.0 nm | $a_1$=4.96 nm | $a_1$=7.08 nm, $r_1$=0.34 | $a_1$=1.59 nm, $r_1$=0.34 |
|  | $g_2$=2.000 | $a_2$=24.0 nm | $a_2$=9.92 nm | $a_2$=14.2 nm, $r_2$=0.34 | $a_2$=3.18 nm, $r_2$=0.34 |
|  | $g_3$=1.000 | $a_3$=12.0 nm | $a_3$=4.96 nm | $a_3$=7.08 nm $r_3$=0.34 | $a_3$=1.59 nm, $r_3$=0.34 |

We present several numerical examples in order to validate the design procedure based on the circuit model approach. Specifically, we include examples for each of the four classes of filters discussed in the previous sections (i.e., low-pass, high-pass, band-pass and band-stop filters). Moreover, we implement three filters with different orders (i.e., $N = 1$, $N = 2$ and $N = 3$) for each of these classes of filters. The numerical results illustrated in Fig. 4 were obtained using the commercial software CST Microwave Studio®. Time domain solver is adopted with automatic hexahedral meshing. We simulate infinite large slabs in the xy-plane by bounding the simulation domain with perfect electric conductors (PEC) in the xz-planes, and perfect magnetic conductors (PMC) in the yz-planes. Two waveguide ports are implemented with y-polarized uniform plane wave to measure the dispersion response using transmission coefficient, i.e., S21.

First, we start by designing three different ($N = 1, 2\ and\ 3$) low-pass filters with cut-off frequency $f_{3dB}$= 400 THz formed by slabs with relative permittivity $\epsilon_d$=10. The thicknesses of the slabs calculated by means of the simple expression (10) are gathered in Table I. The simulated performance is reported in Fig. 4(a), which depicts the frequency response of the transmission coefficient. The simulation results present a reasonable agreement with respect to the response that was expected from the circuital analysis. Specifically, the 3-dB cutoff frequency is 416 THz for the three filters, and the attenuation rate in the cut-off region increases with the order of the filter. Note that the cut-off frequency deviates only a 4% from its prescribed value, which is an excellent result in view of the simplicity of the design equation (10).

Secondly, we transform this low-pass filter into a high-pass filter with cut-off frequency $f_{3dB}$=400 THz. To this end, we make use of plasmonic slabs with plasma frequency 2000 THz, so that the relative permittivity at the cut-off frequency is $\varepsilon_m(\omega_{3dB})$ = -24. Again, the thicknesses of the slabs were calculated from equation (11) and are listed in the third column of Table I. The simulated performance of the filters is shown in Fig. 4(b), where it is observed that the 3-dB cutoff frequency is 405 THz (1.25% deviation from its original value). In consistency with the circuit model, increasing the order of the filters results in a larger transmission above the cut-off frequency, as well as a more acute attenuation below it.

Next, we design a band-pass filter with center frequency $f_0$=500 THz and fractional bandwidth Δ=40% ($f_1$=410 THz and $f_2$=610 THz). As anticipated, the parallel LC circuit units are formed with the combination of plasmonic ($\omega_p = 2\pi \times 5000\ rad/s$, $\epsilon_m(\omega_0)$= -99) and dielectric slabs ($\epsilon_d$=50). The dimensions of both slabs were calculated using equations (12) and (13) and are included in Table I. As shown in Fig. 4(c), the frequency response of the filters is characterized by clear band-pass centered at 506 THz. Naturally, the selectivity of the filter is improved as the order of the filter increases.

However, these results also enable us to identify a constraint in the design of band-pass filters. It is clear from equation (13) that the thicknesses of the overall LC units are inversely proportional to the fractional bandwidth Δ. Therefore, if the targeted Δ is very small, i.e., if we want to design a filter with a very narrow bandwidth, the thickness of each layer would become too large to behave as a lumped element and the response of the device would deviate from the predictions of the circuit model. That is why large absolute values of the relative permittivities were selected for the capacitive and inductive slabs in this example. One way to circumvent this constraint could be to use multi-layered structures, as the 'sandwiched' structure depicted in Fig. 4(c), in order to implement the required lumped elements (which also provides a "symmetric" structure). Another possibility to achieve a narrow band-pass response could be to cascade a high-pass filter with $f_{3dB} = f_1$ and a low-pass filter with $f_{3dB} = f_2$, where $f_1$ is the lower frequency edge and $f_2$ is the higher frequency edge of the passband.

To finalize, Fig. 4(d) represents the transmission properties of our design of a band-stop filter. The center frequency of the filter is set to $f_0$=200 THz, while the prescribed fractional bandwidth is Δ=10% ($f_1$=190 THz and $f_2$=210 THz). In this case, the series LC units are formed by using dielectric ($\epsilon_d$ =10) and plasmonic ($\omega_p = 2\pi \times 500\ rad/s$, $\epsilon_m(\omega_0)$ = -24) materials. The corresponding dimensions of the material strips composing the series LC unit are shown in the fourth column of Table I. Similar to the previous examples, the simulated performance agrees with the expected circuital response, i.e., the stop-band is centered at 197 THz and the attenuation rate increases along with the order of the filter. However, we can again identify a limitation in the design process. Specifically, it is clear from equation (15) that the thickness of the slab is directly proportional to the fractional bandwidth $\Delta$, while it is inversely proportional to the center frequency $\omega_0$. Hence, the required thickness may become unphysically small if the required quality factor (Q) is very large, posing difficulties in the fabrication process.

In conclusion, our results demonstrate that optical metatronics enable the synthesis of desired frequency dispersion via the design of different higher-order optical filters by using the same tools and methods adopted in the design of electronic and microwave filters. The design of the filters is completely carried out by combining lumped circuit elements implemented as electrically thin slabs, and very simple design rules can be derived using metatronic concepts. After the design phase, the performance of the filters has been assessed using full-wave numerical simulations. Our results can be extrapolated to virtually any frequency dispersion synthesis and design procedure of electronic and microwave circuits and, hence, they pave the way to exciting possibilities in the design of composite materials with desired dispersion and highly functional and compact optical filters.

This work was supported in part by the US Air Force Office of Scientific Research (AFOSR) Multidisciplinary University Research Initiative (MURI) grant number FA9550-14-1-0389, in part by the National Natural Science Foundation of China under Contract 61301001 to Y. Li.

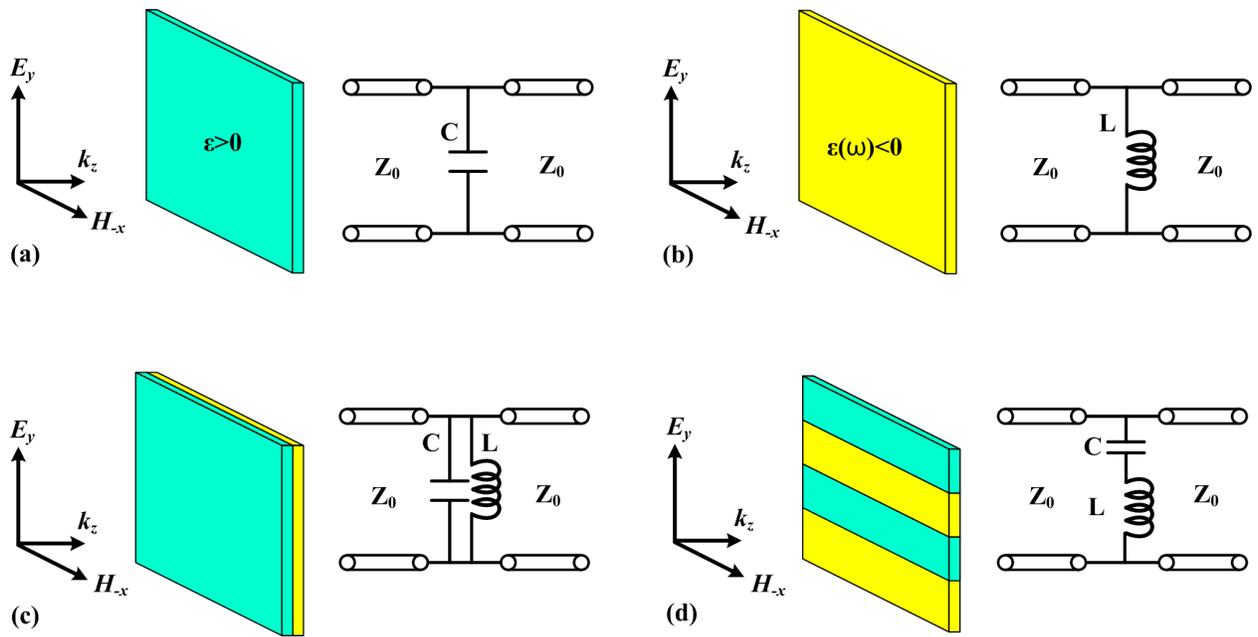

**Fig.1. Sketches and associated circuit diagrams of metatronics-inspired first-order filters:** (a) *Low-pass* filter formed by an electrically thin dielectric (nonmagnetic) material slab. (b) *High-pass* filter formed by an electrically thin plasmonic slab with negative permittivity. (c) *Band-pass* filter formed by two adjacent dielectric and plasmonic slabs (parallel connection). (d) *Band-stop* filter formed by an inhomogeneous slab constructed by alternate dielectric and plasmonic strips (series connection).

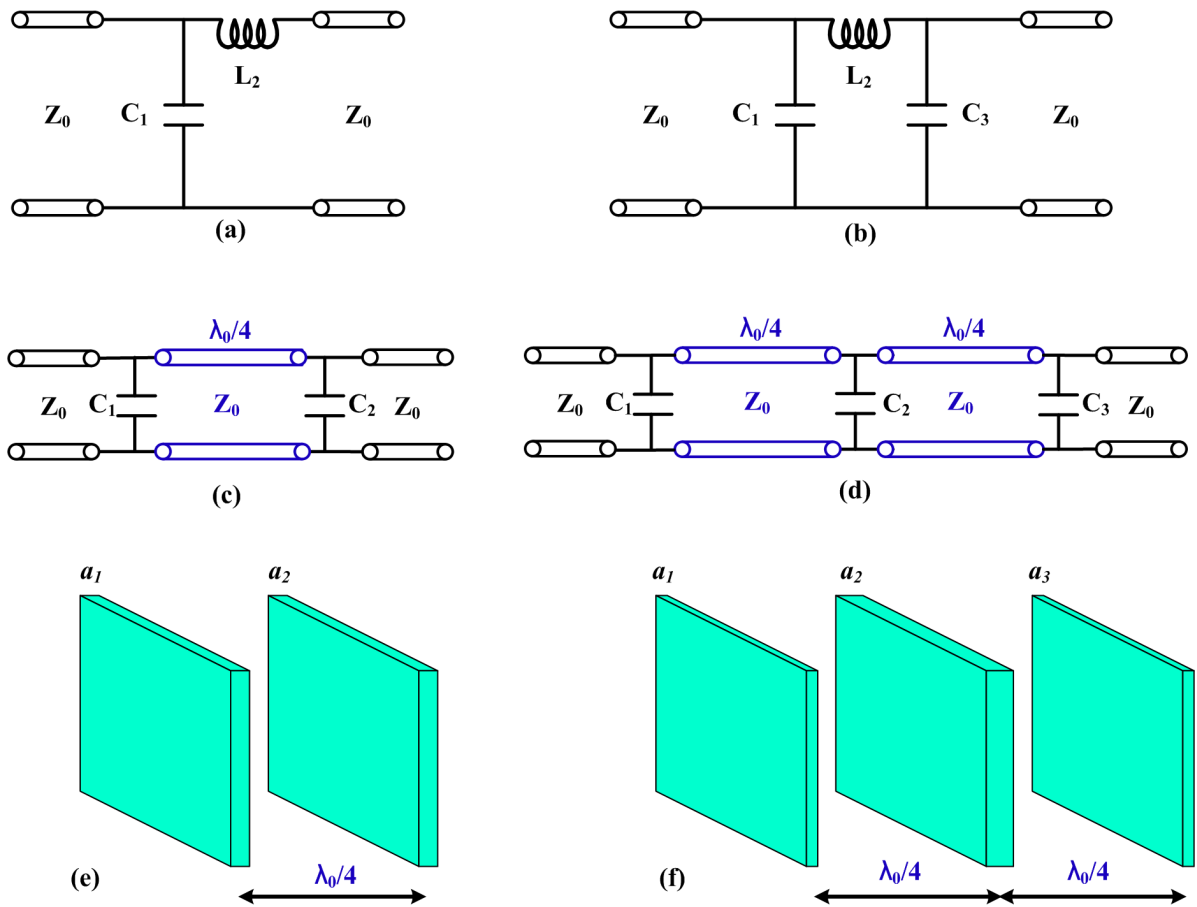

**Fig. 2. Higher-order low-pass filter designs:** Circuit diagrams of (a) $2^{nd}$ order and (b) $3^{rd}$ order filters. Equivalent transmission line models for (c) $2^{nd}$ order and (d) $3^{rd}$ order filters. Sketches of the geometry of the metatronic implementation of (e) $2^{nd}$ order and (f) $3^{rd}$ order filters using thin slabs with a single dielectric material.

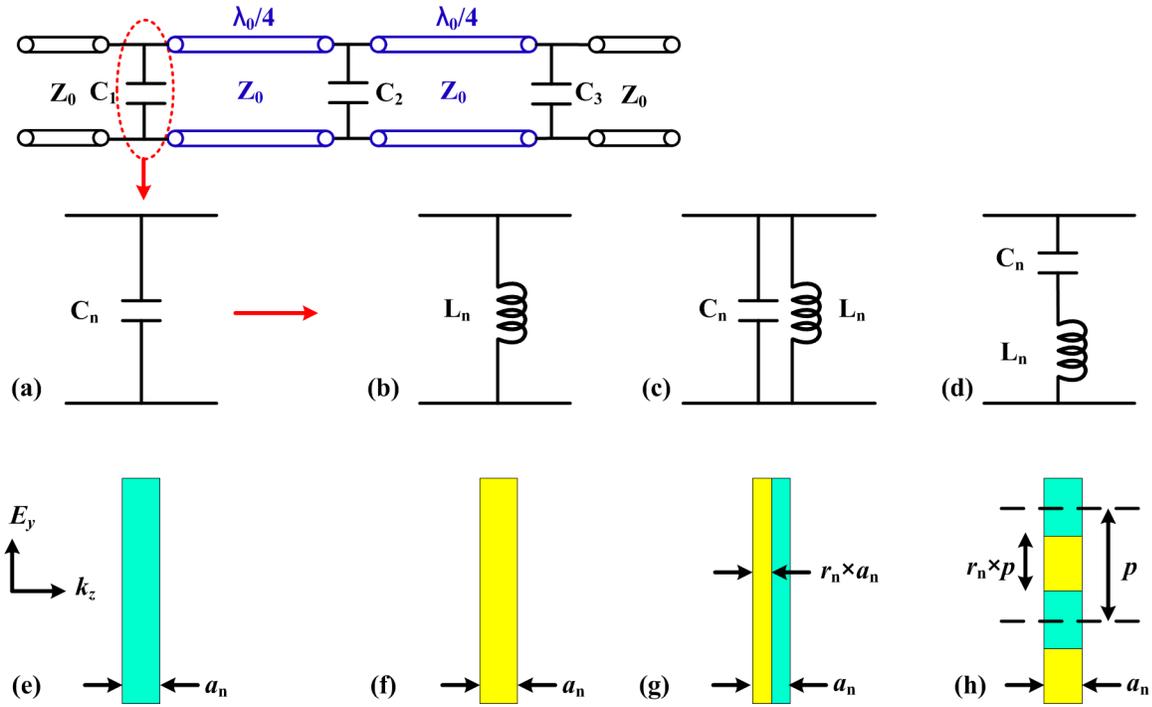

**Fig. 3. Filter transformations:** circuit diagram of the impedance transformation from a (a) low-pass filter design to (b) high-pass, (c) band-pass, and (d) band-stop filters. Sketch of the metatronic implementation of (a) low-pass, (b) high-pass, (c) band-pass and (d) band-stop filters formed by metasurfaces made of only one dielectric and one plasmonic material.

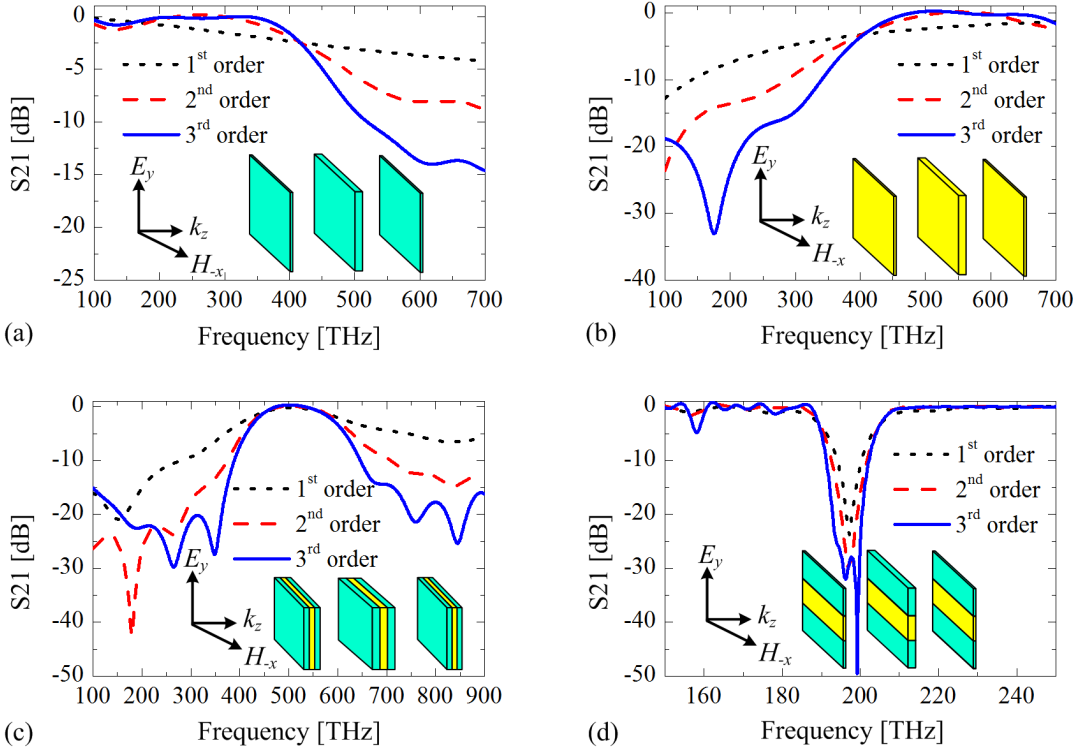

**Fig. 4. Simulated performance of higher-order metatronic optical filters.** Magnitude of the transmission coefficients ($S_{21}$) for 1$^{st}$, 2$^{nd}$, 3$^{rd}$ order optical filters (a) low-pass, (b) high-pass, (c) band-pass, and (d) band-stop frequency response. The geometry of the filters is reported in Table I. Bottom right insets depict a sketch of the geometry of the implementation of a 3$^{rd}$ order filter with layered structures.